\documentclass[twocolumn,fleqn,natbib]{svjour2}

\bibpunct{[}{]}{;}{n}{}{,} % to get "[numbered]" references from natbib
\smartqed  % flush right qed marks, e.g. at end of proof
\usepackage{graphicx}
\journalname{Granular Matter}
\begin{document}

\title{Using NMR to Validate First-Principles Granular Flow Equations\thanks{This work was supported by US National Science Foundation Grant No. CTS-0310006.}}

%\titlerunning{Short form of title}        % if too long for running head

\author{D. Candela \and C. Huan \and K. Facto \and R. Wang \and R. W. Mair \and R. L. Walsworth}

%\authorrunning{Short form of author list} % if too long for running head

\institute{D. Candela \at
Physics Dept., University of Massachusetts, Amherst, MA 01003, USA \\
Tel.: +413-545-3666\\
Fax: +413-545-1691\\
\email{candela@physics.umass.edu}
\and
C. Huan, K. Facto \at
Physics Dept., University of Massachusetts, Amherst, MA 01003, USA            
\and
R. Wang, R. W. Mair, R. L. Walsworth \at
Harvard-Smithsonian Center for Astrophysics, Cambridge, MA 02138, USA
}

\date{Received: date }
% The correct date will be entered by the editor

\maketitle

\begin{abstract}
	Nuclear magnetic resonance (NMR) experiments are described for two granular-flow systems, the vibrofluidized bed and the gas-fluidized bed.
	Using pulsed field gradient, magnetic resonance imaging, and hyperpolarized gas NMR, detailed information is obtained for the density and motions of both grains and interstitial gas.
	For the vibrofluidized bed, the granular temperature profile is measured and compared with a first-principles formulation of granular hydrodynamics.
	For the gas-fluidized bed, dynamic correlations in the grain density are used to measure the bubble velocity and hyperpolarized xenon gas NMR is used to measure the bubble-emulsion exchange rate.
	A goal of these measurements is to verify in earth gravity first-principles theories of granular flows, which then can be used to make concrete predictions for granular flows in reduced gravity.

\end{abstract}

\begin{figure}
\centering
\includegraphics[width=0.45\textwidth]{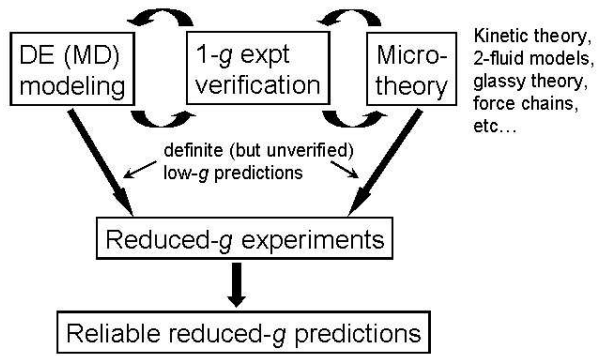}
\caption{\label{boxdiagram}
Routes to reliable predictions of granular-media properties in reduced gravity.
Discrete-element (DE) modeling (sometimes called molecular dynamics (MD) modeling) has the acceleration of gravity $g$ as a direct input.
Similarly, microscopic theories of granular media depend explicitly upon $g$.
Both types of calculation should be tested by physical experiments in earth gravity, before more difficult reduced-gravity experiments are carried out.
}
\end{figure}

\section{Introduction}\label{intro}

	Granular flow and fluidized-bed properties in reduced gravity and interstitial gas pressure will be important for processing, storage, and other applications during lunar and Martian missions.
	However, current engineering practice for granular flows is based on correlations of Earth-gravity data and so cannot be extrapolated with any certainty to reduced gravity.
	An alternative path is to use experiments to validate granular flow equations derived from fundamental physical principles, in which the acceleration of gravity appears as an explicit input parameter.
	To the extent that this is possible, reasonable predictions for reduced gravity granular systems can be made, although it will remain necessary to test the predictions in actual microgravity experiments (Fig.~\ref{boxdiagram}).

	Nuclear magnetic resonance (NMR) enables noninvasive measurements to be made on dense, opaque systems such as granular media.
	We are engaged in a program of experiments to build up detailed pictures of dense, three-dimensional granular systems in Earth gravity using NMR methods combining MRI, PFG-NMR, and hyperpolarized-gas NMR.
	Using these methods we are able to measure density profiles, and grain as well as gas motions over millisecond to second time scales.  We will describe our studies of two granular flow systems: the vibrofluidized bed and the bubbling gas-fluidized bed.

\section{NMR techniques for granular flows}\label{nmr}
	Broadly, two requirements must be met to carry out NMR experiments on a flowing granular medium.
	First, a system must be set up in the desired granular flow state (e.g., bubbling gas-fluidized bed) which also gives a substantial NMR signal for the subsystem to be measured (grains or gas).
	Second, suitable NMR protocols must be devised to extract the desired information such as density, velocity, diffusion, exchange between phases, etc.
	The ways in which these two requirements are met are quite different for grain and gas NMR, yet sometimes it is possible to carry out grain and gas measurements on nearly identical systems.

\begin{figure}
\centering
\includegraphics[width=0.45\textwidth]{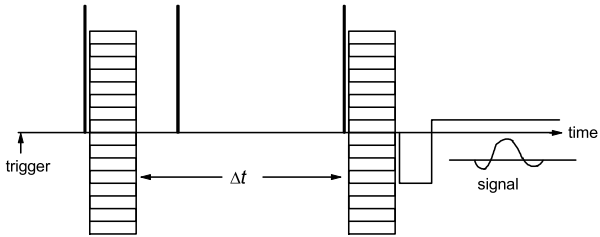}
\caption{\label{sequence}
	Typical NMR pulse sequence used for grain measurements in granular flows.
	The vertical lines indicate radio-frequency pulses, and the rectangles indicate gradient pulses of variable strength.
	Using this type of sequence, it is possible to measure the joint distribution of grain motions during the interval $\Delta t$ with $\mu$m resolution, and grain positions with sub-mm resolution.
	The entire sequence can be completed in as little as 2~ms, although it must be repeated many times to build up the distribution information.}
\end{figure}

\begin{figure}
\centering
\includegraphics[width=0.45\textwidth]{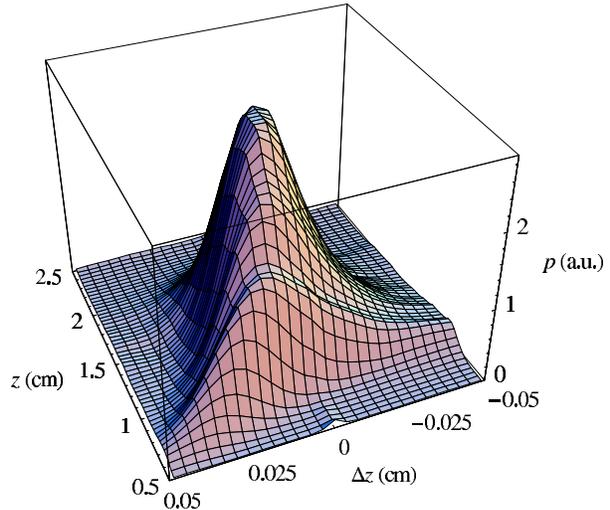}
\caption{\label{pdzz}
	Joint distribution $p(\Delta z,z)$ of vertical displacements $\Delta z$ in $\Delta t = 1.38$~ms and heights $z$, for the grains in a vibrofluidized bed of mustard seeds \cite{huan04}.
	Near the bottom of the bed (small $z$) the distribution is highly skewed due to collisions with the vibrating bottom wall, while at large $z$ the distribution is nearly Guassian.}
\end{figure}

\subsection{Grain NMR techniques}\label{graintech}
	Due to the rapid molecular tumbling, fluids give NMR signals much more suitable for imaging-type studies than do solids~\cite{abragam61}.
	Botanical seeds, which typically contain liquid oil that gives a strong NMR signal, have been one popular type of sample for granular-media studies~\cite{caprihan97, knight96}.
	To allow NMR studies over a wider range of grain sizes, small amounts of organic liquids can be adsorbed onto substrates such as porous catalyst particles~\cite{savelsberg02}.
	In the experiments described here, both types of grains were employed.

	For grain-phase studies, we typically use NMR pulse sequences that can encode both the motion of the grains and their positions (Fig.~\ref{sequence}).
	Using strong gradient pulses (up to 1000~G/cm), it is possible to resolve grain motion over times of one millisecond or less.
	In this type of study, the individual grains are not imaged.
	Rather, the joint probability distribution over all grains is measured as a function of grain position and grain displacement in a specified time interval~\cite{callaghan91}.
	Figure~\ref{pdzz} shows an example of this type of data.
	Note that for our studies of systems with continuous, rapid grain movement both the NMR techniques used and the type of data obtained are quite different from ealier studies of quasistatic flows~\cite{knight96}.

\begin{figure*}
\centering
\includegraphics[width=0.75\textwidth]{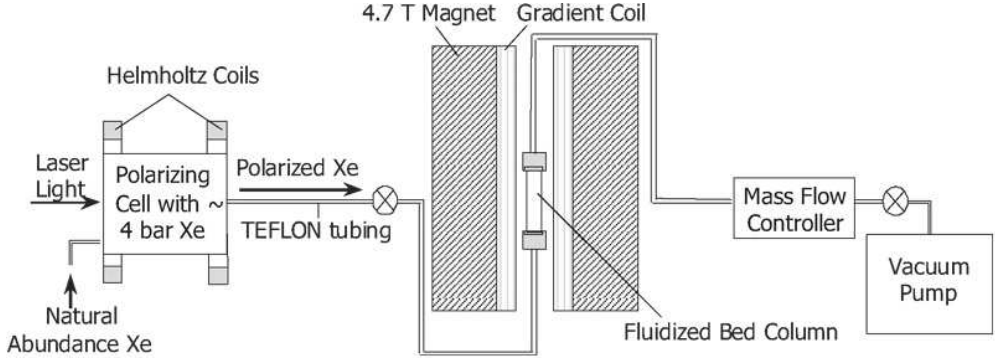}
\caption{\label{apparatus}
Simplified diagram of apparatus for using hyperpolarized xenon to study gas motion in a fluidized granular bed.}
\end{figure*}

\subsection{Gas NMR techniques}\label{gastech}
	Under equilibrium conditions, the NMR signal from gasses is very weak (due to their low density), often making gas-phase NMR studies difficult.
	By using hyperpolarized gas, which has been shown to be practical for $^3$He and $^{129}$Xe, the NMR signal can be increased by a factor of order $10^4$~\cite{walker97}.
	Figure~\ref{apparatus} shows a simplified diagram of the apparatus we use to study the gas motion in a gas-fluidized granular bed.
	In this apparatus, xenon gas is polarized by spin exchange with rubidium vapor in a laser optical pumping cell.
	The hyperpolarized xenon is then used as the fluidizing gas for a small granular bed located within the NMR magnet.

\section{The vibrofluidized bed: Granular hydrodynamics}\label{vibro}
	The first granular system to which we have applied these techniques is the vibrofluidized bed.
	Mechanical vibration is a fundamental and widely used technique for transforming granular media from solid-like to fluid-like states.
	In addition, vibration is often used as an aid for other fluidization methods such as gas flow or gravity-driven flow.
	Questions one would like to be able to answer from basic physical principles include: What amplitude and frequency of vibration are required for fluidization?  What are average grain density and grain speeds as functions of height within the bed?

	One theoretical approach has been to treat the individual grains like molecules in a (pseudo) fluid, and to bring to bear methods of statistical physics such as kinetic theory and hydrodynamics~\cite{grossman97, haff83, jenkins83, vannoije98, sela98}.
	This approach is most tractable (and more likely to be valid) in the ``rapid granular flow'' regime, in which the transient two-body collisions between the grains dominate over longer-lived and many-particle interactions~\cite{campbell90}.
	For a vibrofluidized bed, this regime occurs for strong, high frequency driving of relatively shallow beds~\cite{huan04}.

\begin{figure}
\centering
\includegraphics[width=0.45\textwidth]{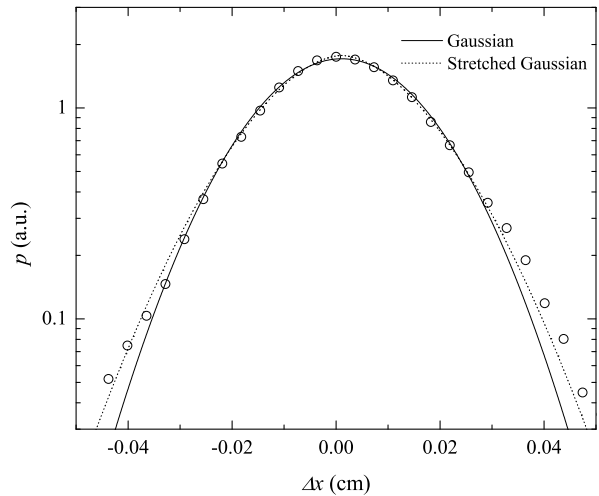}
\caption{\label{gauss}
	Distribution of horizontal grain displacements in a vibrofluidized bed under the same conditions as Fig.~\ref{pdzz}.
	The NMR observation time $\Delta t$ was 1.38~ms, considerably shorter than the mean grain collision time $\approx 6$~ms \cite{huan04}.
	Therefore, the measured displacement distribution should give a good approximation to the grain \emph{velocity} distribution.
	The curves show fits to a Gaussian function with and without power-law stretching of the exponent.
	The distribution deviates only slightly from an unstretched Gaussian function, which would describe the distribution of velocities of molecules in a molecular fluid.}
\end{figure}

\begin{figure}
\centering
\includegraphics[width=0.45\textwidth]{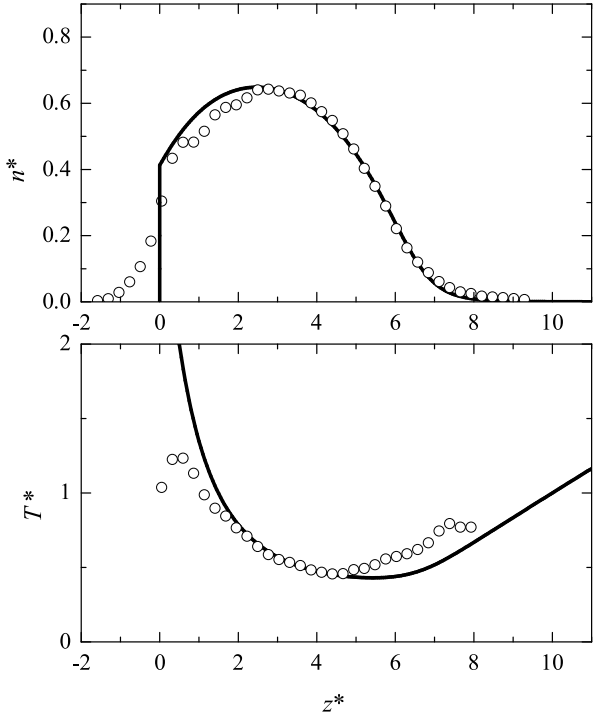}
\caption{\label{nstar}
	Density $n^* = nd^3$(top) and granular temperature $T^*=T/mgd$ (bottom) as functions of height $z^*=z/d$, for a vibrofluidized bed under the same conditions as Fig.~\ref{pdzz} (the starred variables have been nondimensionalized by the grain diameter and mass $d, m$ and the acceleration of gravity $g$).
	The symbols show NMR measurements, while the curves show a fit to the granular-hydrodamics theory of Ref.~\cite{garzo99}.
	A notable feature (in both the experiment and the theory) is a ``temperature inversion''  --- even though the energy flow is strictly upwards, the temperature falls with increasing height $z^*$ and then rises again at large heights.
	At heights above the temperature minimum, the flow of granular ``heat'' is from colder to hotter temperatures, an effect that is only possible in highly out-of-equilibrium systems.
}
\end{figure}

	Several earlier studies had used video techniques to measure grain motion statistics in quasi-two dimensional vibrofluized beds~\cite{clement91, losert99, rericha01, rouyer00, warr95}.
	Some of these two-dimensional studies found strong deviations of the grain velocity distribution from the nearly-Boltzmann distribution that might be expected from the statistical models.
	Using NMR, we have been able to noninvasivly probe the grain velocity distribution in a three-dimensional bed.  For the conditions of our experiments we find that the grain velocity distribution is close to the Boltzmann distribution (Fig.~\ref{gauss}), validating a statistical approach.
	Beyond this, we are able to make a detailed joint fit of the measured density and granular temperature profiles, to a first-principles calculation~\cite{garzo99} of granular hydrodynamic coefficients (Fig.~\ref{nstar}).

	The hydrodynamic equations used to generate the theory curves in Fig.~\ref{nstar} explicitly reference gravity (using a Bernoulli relation, see Ref.~\cite{huan04} for details).
	After validating the hydrodynamic theory in Earth gravity by a series of fits like that shown in Fig.~\ref{nstar}, it is reasonable and easy to extend the theory to reduced gravity by recalculating it with reduced $g$.
	The output of such a calculation could be a prediction for the density profile in a granular bed vibrofluidized in lunar or Martian gravity.
	This is a (perhaps rather simplistic) example of the program outlined in Fig.~\ref{boxdiagram}.

\section{The gas-fluidized bed: Grain and gas NMR}\label{gasfluid}
	Gas-fluidized granular beds are widely used in industry for catalytic cracking, efficient combustion, and many other large-scale processes~\cite{geldart86,kunii91,kwauk01}
	It is reasonable to imagine that gas fluidization (or pneumatic transport of granular material, which is closely related) would play a prominent role in in-situ resource utilization on the moon or Mars, yet the behavior of a gas-fluidized bed in reduced gravity is almost completely unknown at present.

	To promote mixing, gas-fluidized beds are typically used in the bubbling regime, in which large bubble-like voids devoid of grains rise through the bed at a velocity much larger than the mean gas velocity.
	From a theoretical standpoint the bubbling gas-fluidized bed is extremely complex, with significant motion over a wide range of length scales (from grain to bubble) and fluid dynamics with self-consistently moving boundary conditions.
	Microscopically-motivated theories of fluidized-bed behavior have been formulated (see, for example Ref.~\cite{jackson00}) yet to achieve closure these theories must make various unverified assumptions.
	For example, in a ``two fluid'' model it might be assumed that the grains and gas behave like two interpenetrating Newtonian fluids~\cite{jackson00}.
	Yet, at present it is not established theoretically or experimentally that a \emph{dense} granular medium is ever well described as a Newtonian fluid.

	In common with other granular media, the gas-fluidized bed is difficult to probe experimentally in a non-invasive manner.
	To better understand the dynamics of the bubbles and their interaction with the dense (``emulsion'') phase of the bed, many techniques have been tried including x-ray photography and tracer-gas injection~\cite{geldart86}.
	Here we show preliminary data for two projects we are carrying out using NMR to probe the bubbles.
	Neither project has reached the stage of detailed agreement with first-principles theory, as shown above for the vibrofluidized bed.
	Nevertheless some quantitative connections with model predictions are made.

\begin{figure*}
\centering
\includegraphics[width=0.75\textwidth]{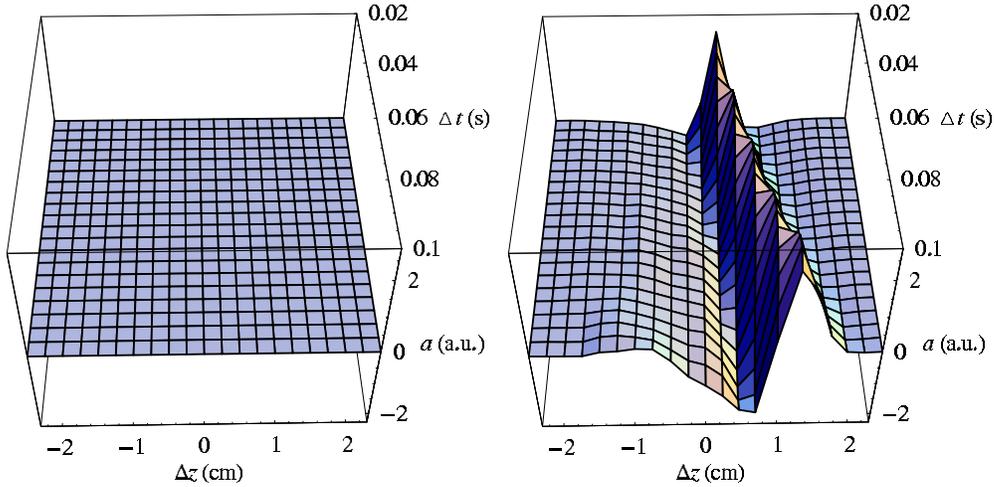}
\caption{\label{corr}
	Preliminary measurement of correlations between fluctuations in the density of a gas-fluidized bed of 75-106~$\mu$m diameter alumina particles, measured using NMR.
	A small quantity of dodecane is adsorbed onto the porous particles, to give NMR sensitivity.
	The correlations are plotted as a function of the time shift $\Delta t$ and the height shift $\Delta z$.
	For the left plot, there is no gas flow through the bed and hence no fluctuations.
	For the right plot, there is a gas flow sufficient to put the bed into the bubbling regime (superficial gas velocity $v_g=1.4$~cm/sec).
	In this case there are large density fluctuations, due to the bubbles, and the bubble velocity $v_b=12.4$~cm/s can be measured directly from the slope of the ridge in the correlation plot.}
\end{figure*}

\subsection{Measurement of bubble velocity using grain NMR}\label{bubgrain}
	Simple and basic questions for the bubbling gas-fluidized bed are: what is the velocity $v_b$ of the rising bubbles, and how is the velocity related to the bubble diameter $d$, gas flow rate, and other bed parameters?
	In analogy with gas bubbles in a liquid (or by dimensional analysis), it can be argued $v_b \sim (dg)^{1/2}$, and such a relation has been proposed for the gas-fluidized granular bed with a phenomenological prefactor~\cite{davidson63}.
	Yet it is not easy to measure the mean velocity of bubbles rising in the interior of a chaotically bubbling gas-fluidized bed.

	We have devised a new method to measure the mean velocity and other characteristics of the bubbles, using NMR on the grains.
	A pulse sequence (different from that shown in Fig.~\ref{sequence}) is used to measure the bed density profile twice, separated by a time interval $\Delta t$, and the experiment is repeated many times.
	Bubbles in the bed appear as localized, downward fluctuations in the bed density.
	When the cross correlation between the fluctuations in the two time-separated density measurements are plotted (Fig.~\ref{corr}) a ridge appears corresponding to fluctuations that propagate at a definite velocity.
	The mean velocity of the bubbles can be accurately measured from the slope of the ridge, while the width and shape of the ridge presumably indirectly reflect the size and shape of the bubbles.
	This analysis has not yet been completed, but it is already clear that the bubble velocity and size (from the slope and width of the ridge) are approximately consistent with the excess gas flow over minimum fluidization flow.
	Similarly, the bubble size and velocity from the graph are approximately consistent with the $d$ vs. $v_b$ correlation proposed in Ref.~\cite{davidson63}.

\begin{figure}
\centering
\includegraphics[width=0.45\textwidth]{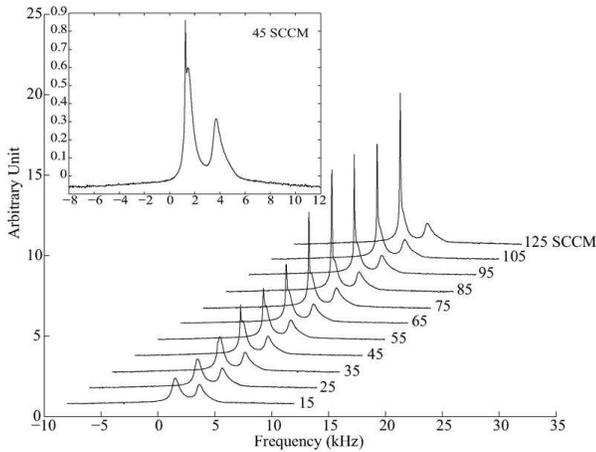}
\caption{\label{spectra}
	$^{129}$Xe spectra for an alumina-particle bed fluidized by hyperpolarized xenon gas, measured at 11 different gas flow rates ranging from 15 to 125 sccm.
	Three peaks may be distinguished: a narrow peak due to gas within the bubbles, a broader peak at nearly the same frequency due to gas within the interstitial volume of the emulsion (dense granular) phase, and a second broad peak shifted to higher frequency due to gas adsorbed on the porous alumina particles.
	As the flow rate is increased, the fraction of gas in the bubbles also clearly increases.}
\end{figure}

\subsection{Measurement of bubble-emulsion exchange using hyperpolarized gas NMR}\label{bubgas}

	Using the setup shown in Fig.~\ref{apparatus}, we are able to perform NMR experiments directly on the gas within the fluidized bed.
	Significantly, we have used the same porous alumina particles and the same bed diameter as for the grain-NMR experiments summarized above.
	Therefore, we can correlate grain and gas NMR information in essentially identical fluidization conditions (as verified by visual measurements of the bed-height hysteresis curve, which is extremely sensitive to the fluidization state).

\begin{figure}
\centering
\includegraphics[width=0.45\textwidth]{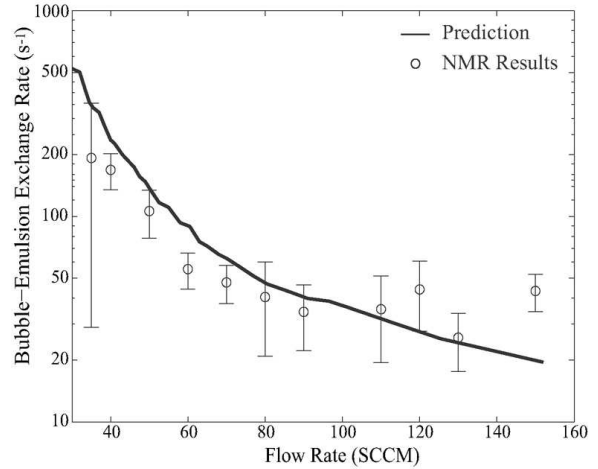}
\caption{\label{exchange}
	Preliminary measurement of bubble-emulsion exchange rate as a function of gas flow rate for a bed of alumina particles fluidized by hyperpolarized xenon gas.
	The exchange rate gives the rate at which gas within the bubbles interchanges with gas within the dense granular phase, and is a key parameter for engineering applications of fluidized beds.
	The solid curve shows the exchange rate calculated from the measured bed expansion along with the phenomenological theory from Ref.~\cite{davidson63}, while the points with error bars show a direct measurement of the exchange rate using $^{129}$Xe NMR.}
\end{figure}

	The $^{129}$Xe NMR frequency is sensitive to the physical environment of the gas molecule, which makes it possible to distinguish spectroscopically between gas in the bubbles, gas between the grains, and gas adsorbed onto the grains (Fig.~\ref{spectra}).
	Using a suitably tailored NMR pulse sequence \cite{wang05}, we are able to selectively eliminate the signal from gas outside the bubbles and then observe this signal re-grow due to gas motion in the bed.
	In this way an important engineering parameter for gas-fluidized beds, the bubble-emulsion exchange rate, can be directly and noninvasively measured (Fig.~\ref{exchange}).
	The exchange rate $K$ is defined by the following equation ~\cite{kunii91}:
\[
\frac{1}{V_b}\frac{dN_b}{dt} = -K(C_b-C_e)
\]
where $N_b$ is the quantity of a tracer gas within a bubble of volume $V_b$, and $C_b, C_e$ are the concentrations of the tracer in the bubble and emulsion phase respectively.
	For our experiments the hyperpolarized xenon gas serves as a completely neutral tracer, and its concentration in the various phases can be measured accurately and noninvasively using NMR.

\section{Conclusions}
	Using NMR methods, it is possible to measure the densities and motion of both grains and gas in continuously excited granular systems such as the vibrofluidized and gas-fluidized beds.
	Quantitative results can be obtained for many parameters including the grain density profile, granular temperature (mean random grain kinetic energy), grain bubble velocity and size, and gas bubble-emulsion exchange rate.

	Where first-principles physical theories exist for the granular flow state, e.g., the highly excited vibrofluidized bed, these theories can be directly tested.
	Conversely, for more complex systems like the gas-fluidized bed theoretical models must make more ad-hoc assumptions (for example that the bubbles behave approximately as bubbles in a fluid).
	In this case non-invasive NMR measurements offer possibilities to directly test the results of such assumptions (e.g., predictions of the bubble velocity-size relationship and bubble-emulsion exchange rate).

	In both cases, a key goal is to understand the granular system sufficiently well to enable extrapolations to as yet unexplored regions of parameter space, such as reduced-gravity operations.
	The difficulty and risk of actual reduced-gravity experiments render well-supported predictions for reduced gravity especially valuable.

\bibliographystyle{spbasic}
\bibliography{nasaproc,granflow}

\end{document}